# Flat Dielectric Response in $3BaO$-$3TiO_2$-$B_2O_3$ Glasses

Rahul Vaish and K.B.R. Varma*

Materials Research Centre,

Indian Institute of Science,

Bangalore 560 012,

India.

1

*Corresponding Author; E-Mail : kbrvarma@mrc.iisc.ernet.in;

FAX: 91-80-23600683; Tel. No: 91-80-22932914

**Abstract:**

X-ray powder diffraction (XRD) along with differential thermal analysis (DTA) carried out on the as-quenched samples in the $3BaO$-$3TiO_2$-$B_2O_3$ system confirmed their amorphous and glassy nature, respectively. The dielectric constants in the 1 kHz-1 MHz frequency range were measured as a function of temperature (323-748 K). The dielectric constant and loss were found to be frequency independent in the 323-473 K temperature range. The dielectric behavior was consistent with the universal dielectric response (UDR). The temperature coefficient of dielectric constant was estimated using Havinga's formula and found to be 16 ppm.$K^{-1}$. The electrical relaxation was rationalized using the electric modulus formalism. The dielectric constant and loss were 17 ± 0.5 and 0.005 ±0.001, respectively at 323 K in the 1 kHz-1 MHz frequency range which may be of considerable interest to capacitor industry.

**1. Introduction:**

Borate-based polar single crystals have been of increasing importance owing to their promising physical properties which could be exploited for non-linear optical, pyroelectric, piezoelectric and surface acoustic wave (SAW) device applications. Single crystals such as $LiB_3O_5$, $Li_2B_4O_7$, $CsLiB_6O_{10}$, $BiB_3O_6$ were reported to possess interesting physical properties [1-3]. $Ba_3Ti_3O_6(BO_3)_2$ (BTBO) single crystals were found to exhibit excellent non-linear optical properties [4]. These materials could also be used as high dielectric constant glasses due to the high polarizability associated with $Ti^{+4}$ ion. We have been making systematic attempts to fabricate glasses and glass-ceramics in the $BaO$-$TiO_2$-$B_2O_3$ system to visualize their physical properties. The objectives have been to be grow nano/micrometer sized crystallites in a transparent glass matrix of the same composition which would ensure transparency. These transparent glass nano/microcrystal composites would be particularly interesting from their optical, electro-optic and non-linear optical characteristics view point. Indeed the size and morphology of the crystallites apart from their connectivity (which is an important parameter for demonstrating ferroelectric properties) can be engineered by choosing appropriate heat-treatment temperatures. In the present investigations the compositions in the above system were chosen such that one eventually obtains $Ba_3Ti_3O_6(BO_3)_2$ (BTBO) crystalline phase on crystallization [5]. Though, $BaO$-$TiO_2$-$B_2O_3$ glass system was investigated for its structural, optical and crystallization behavior [6], the composition which would yield the BTBO crystalline phase on heat-treatment has not been reported. Therefore, to begin with, glasses in the composition $3BaO$-$3TiO_2$-$B_2O_3$ have been investigated for their dielectric and electrical relaxation properties (as these are related to their electro-optic and non-linear optical properties ) over the range of temperatures and frequencies that are normally of interest in the applications of these materials. The details of which are reported in this article.

**2. Experimental:**

Transparent glasses in the composition $3BaO-3TiO_2-B_2O_3$ (in molar ratio) were fabricated via the conventional melt-quenching technique. For this, $BaCO_3$, $TiO_2$ and $H_3BO_3$ were mixed and melted in a platinum crucible at 1473K for 1h. The batch weight was 10 gm. Melts were quenched by pouring on a steel plate that was maintained at 423K and pressed with another plate to obtain 1-1.5 mm thick glass plates. All these samples were annealed at 773 K (5 h) which is well below the glass transition temperature. X-ray powder diffraction (Philips PW1050/37, Cu *K*a radiation) study was performed on the as-quenched powdered samples at room temperature to confirm their amorphous nature. The DTA (TA Instruments SDTQ 600) runs were carried out in the 573K –1173K temperature range. The glasses under study were heated at temperatures corresponding to their crystallization temperatures (as predetermined by the DTA studies) for a few hours and subjected to X-ray diffraction studies.

The capacitance and dielectric loss (D) measurements on the as-quenched (annealed) polished glass plates that were gold sputtered were done using impedance gain phase analyzer (HP 4194 A) in the 1 kHz-1 MHz frequency range with a signal strength of 0.5 $V_{rms}$ at various temperatures (332–748 K). Thin silver leads were bonded to the sample using silver epoxy. Based on these data, the dielectric constant was evaluated by taking the dimensions and electrode geometry of the sample into account.

## 3. Results and discussion:

The DTA trace that was obtained for the as-quenched glass plates of the thickness 1mm at a heating rate of 10 K/min is shown in Fig. 1. An endotherm followed by an exotherm are observed for the as-quenched samples understudy. The endotherm at 827 K corresponds to the glass-transition temperature. A subsequent exotherm that is encountered around 920 K is attributed to the crystallization of the $3BaO\text{-}3TiO_2\text{-}B_2O_3$ glasses. The X-ray powder diffraction (XRD) pattern recorded for the as-quenched glasses confirms their amorphous nature (Fig. 2). The variation of the dielectric constant ($\varepsilon_r^{'}$) with frequency (1 kHz-1 MHz) of measurement for the glasses at different temperatures (323 K-523 K) is shown in the inset of Fig. 3. It is seen that the dielectric constant has weak dependence on frequency over this range upto the 473 K. For the present glasses, the average value obtained for the dielectric constant is 17±0.5 at 323K in the frequency range under study. The dielectric constant increases marginally with the increase in temperature at all the frequencies under study. At low frequencies the $\varepsilon_r^{'}$ increases by about two which may be ascribed to the mobile ion polarization combined with electrode polarization. The dielectric behavior of the present glass system is rationalized by invoking Jonscher's universal law [7].

$$(\varepsilon_r^{'} - \varepsilon_\infty) \text{ a } f^{\,n-1} \qquad (1)$$

after taking logarithm of the above equation;

$$\log (\varepsilon_r^{'} - \varepsilon_\infty) \text{ a } (n-1) \log f \qquad (2)$$

The plots of log $\varepsilon_r^{'}$ versus log $f$ along with linear fit (solid line) to the above equation at various temperatures are depicted in the Fig. 3, which demonstrate the validity of the Jonscher's law. It is to be noted that the value of $\varepsilon_\infty$ is not subtracted from the value of $\varepsilon_r^{'}$ during fitting the experimental data as there is no need of subtracting $\varepsilon_\infty$ for the materials of the present kind in which the dielectric response is flat [8]. By fitting the data, the value of *n* is found to be 0.99 ±0.01 in the 323-473K temperature range. As per Kramers-Kronig relations, frequency independent behavior of the dielectric constant ($n \rightarrow 1$) indicates frequency independent low dielectric loss behavior [7]. This corresponds to the limiting form of lattice loss. This response is referred to as "flat loss" corresponding to a low loss behavior, almost independent of the frequency. As per Jonscher's relation based on energy criterion [7],

$$\frac{\varepsilon^{''}}{\left(\varepsilon^{'} - \varepsilon_\infty\right)} = \cot\left(n\pi / 2\right) \quad (3)$$

It indicates that the ratio of energy lost per radian by energy stored is cot (*n*p/2) which implies a low energy loss for a high value of *n* and a high loss for a small value of *n*. It represents a pure capacitance for *n* = 1 (stored energy) and a pure conductance for *n* = 0 (lost energy). The present glass system exhibits low loss state (pure capacitance) since *n* is close to 1 in the 323-473K temperature range. The value of *n* is found to be 0.97 ±0.01 at 523K which indicates that the value for *n* decreases with increase in temperature. Fig. 4 shows the plots of dielectric loss against frequency at various temperatures. Frequency independent (flat) and low loss behaviour was observed in the 323-473K temperature range. The dielectric loss (D) at 323K is 0.005±0.001 at all the frequencies understudy.

The temperature dependent dielectric constant behavior of the as-quenched glasses was also investigated in the 323-673K temperature range over the frequency 1 kHz-1 MHz range (Fig. 5). It is noted from the figure that, at each frequency, the increment in dielectric constant is approximately linear from 323K to about 575K and subsequently becomes non-linear as the temperature increases. It is also observed that the non-linearity that sets in at a lower temperature at low frequencies (1 kHz) shifts towards higher temperatures as the frequency is increased (1 MHz). The temperature dependence of the dielectric constant at different frequencies is rationalized using Havinga's model [8]. According to this model, the temperature dependence of the dielectric constant at constant pressure, is of the form,

$$\frac{1}{(\varepsilon_r^{'}-1)(\varepsilon_r^{'}+2)}\left(\frac{\partial \varepsilon_r^{'}}{\partial T}\right)_P = A+B+C \tag{4}$$

The temperature variation of the dielectric constant for the perfect solid dielectrics depends on three factors ($A$, $B$ & $C$) that are related to the thermal expansion and polarizability of the material. In the above relation "*A*" represents the decrease in the number of polarizable particles per unit volume as the temperature increases and has a direct effect on the volume expansion. "*B*" signifies the increase in the polarizability of a constant number of particles as the volume increases and "*C*" denotes the change in polarizability due to temperature changes at a constant volume. In the present study, the temperature coefficient of dielectric constant $\left[(\varepsilon_r^{'}-1)(\varepsilon_r^{'}+2)\right]^{-1}\left(\partial\varepsilon_r^{'}/\partial T\right)_P$ for the linear region of dielectric constant (Fig. 5) in the 323-575K temperature range was estimated. The value of the coefficient is almost independent of the frequency in the 323-575K temperature range. The average value is 16±1 ppm.K$^{-1}$.

In order to have further insight into the low dielectric loss behavior of these glasses, its frequency-dependent electrical conductivity at various temperatures is studied. Conductivity, at different frequencies and temperatures, was calculated by using the dielectric data as per the following formula;

$$\sigma_{\omega} = \omega \varepsilon_{o} D \varepsilon_{r}^{'} \quad (5)$$

where $\sigma_{\omega}$ is the conductivity at an angular frequency, ? (=2p*f*). The frequency dependence of the conductivity at various temperatures is shown in Fig. 6. The phenomenon of the conductivity dispersion in solids is generally analyzed using Jonscher's law [7];

$$\sigma_{\omega} = \sigma_{DC} + \sigma_{AC} = \sigma_{DC} + Af^{n} \quad (6)$$

where $\sigma_{DC}$ is the DC conductivity and *A* is temperature dependent constant and *n* is the power law exponent which generally varies between 0 to 1. For a wide variety of ionic conductors, it has been observed that at sufficiently low temperatures and/or high frequencies, the conductivity varies linearly or nearly linearly with frequency. Such a behavior is usually known as near constant loss (NCL) behavior which has been reported to be a universal feature of many materials [9-11].

To accommodate the NCL contribution, a linear term of frequency has been introduced in universal dielectric response (Eq. 6) to describe the ac conductivity of the glasses;

$$\sigma_{\omega} = \sigma_{DC} + A.f^{n} + B.f^{1.0} \quad (7)$$

The conductivity in the 323-473K temperature range increases with frequency as a function of $B.f^{1.0}$ suggesting the domination of NCL contribution to the ac conductivity in the frequency range under study. The absence of frequency independent conductivity ($\sigma_{DC}$) region within the measured frequency window indicates lack of long-range ionic diffusion in the 323-473K temperature range. At high temperatures (573-673 K), the conductivity plots could be parameterized using the Eq. 7. At these temperatures, the frequency window of NCL behavior decreases with increase in temperature.

Inorder to analyze the dielectric data further, the electric modulus formalism was invoked. The use of electric modulus approach helps in gaining an insight into the bulk response of materials. Therefore, this approach could effectively be employed to separate out electrode effects. This would also facilitate to circumvent the problems caused by electrical conduction which might mask the dielectric relaxation processes. The complex electric modulus ($M^*$) is defined in terms of the complex dielectric constant ($e^*$) and is represented as [12]:

$$M^* = (e^*)^{-1} \qquad (8)$$

$$M' + iM'' = \frac{\varepsilon_r'}{(\varepsilon_r')^2 + (\varepsilon_r'')^2} + i\frac{\varepsilon_r''}{(\varepsilon_r')^2 + (\varepsilon_r'')^2} \qquad (9)$$

where $M'$, $M''$ and , $\varepsilon_r'$, $\varepsilon_r''$ are the real and imaginary parts of the electric modulus and dielectric constants, respectively. The real and imaginary parts of the modulus at different temperatures are calculated using Eq. 9 for the glasses under study and are depicted in Figs. 7 and 8, respectively. One would conclude from Fig. 7 that at low frequencies, $M'$ approaches zero at all the temperatures under study suggesting the suppression of the electrode polarization. $M'$ reaches a maximum value corresponding to $M_\infty = (\varepsilon_\infty)^{-1}$ due to the relaxation process. It is also observed that the value of $M_\infty$ decreases with the increase in temperature. The imaginary part of the electric modulus (Fig. 8) is indicative of the energy loss under electric field. The $M''$ peak shifts to higher frequencies with increasing temperature. This evidently suggests the involvement of temperature dependent relaxation processes in the present glasses.

The frequency regime that is below the $M^{''}$ peak position indicates the range in which the ions drift to long distances. In the frequency range which is above that of the peak, the ions are spatially confined to potential wells and free to move within the wells. The frequency range where the peak occurs is suggestive of the transition from long-range to short-range mobility. The electric modulus could be expressed as the Fourier transform of a relaxation function $\phi(t)$ [13, 14]:

$$M^* = M_\infty \left[ 1 - \int_0^\infty \exp(-\omega t) \left( -\frac{d\phi}{dt} \right) dt \right] \tag{10}$$

where the function $\phi(t)$ is the time evolution of the electric field within the materials and is usually taken as the Kohlrausch-Williams-Watts (KWW) function [13, 14]:

$$\phi(t) = \exp\left[ -\left( t/\tau_m \right)^{\beta} \right] \tag{11}$$

where $t_m$ is the conductivity relaxation time and the exponent $\beta$ (0 1] indicates the deviation from Debye type relaxation. The value of $\beta$ could be determined by fitting the experimental data in the above equations. But it is desirable to reduce the number of adjustable parameters while fitting the experimental data. Keeping this point in view, the electric modulus behavior of the present glass system is rationalized by invoking modified KWW function suggested by Bergman. The imaginary part of the electric modulus ($M^{''}$) could be defined as [15]:

$$M'' = \frac{M^{''}_{\max}}{(1-\beta) + \frac{\beta}{1+\beta}\left[ \beta(\omega_{\max}/\omega) + (\omega/\omega_{\max})^{\beta} \right]} \tag{12}$$

where $M''_{\max}$ is the peak value of the $M''$ and ? $_{max}$ is the corresponding frequency. The above equation (Eq. 12) could effectively be described for $\beta \geq 0.4$. Theoretical fits of Eq. 12 to the experimental data are illustrated in Fig. 8 as the solid lines. The experimental data are well fitted to this model except in the high frequency regime. From the fitting of $M''$ versus frequency plots, the value of $\beta$ was determined. From the fitting of $M''$ versus frequency plots, the value of $\beta$ is found to be 0.85 ±0.02 in the 648-723 K temperature range indicating the narrow distribution of relaxation times due to low concentration of conducting species in the present glasses. It also reflects the local structural environment of the glasses in homogeneous glasses [16].

It is of interest to investigate into the transport mechanism in present the glasses. Therefore, the DC conductivity at different temperatures ($\sigma_{DC}(T)$), was calculated from the electric modulus data. The DC conductivity could be extracted using the expression [12]:

$$\sigma_{DC}(T) = \frac{\varepsilon_o}{M_{\infty}(T) * \tau_m(T)} \left[ \frac{\beta}{\Gamma\left( 1/\beta \right)} \right] \tag{13}$$

where $\varepsilon_o$ is the free space dielectric constant, $M_{\infty}(T)$ is the reciprocal of high frequency dielectric constant and $\tau_m(T)$ is the temperature dependent relaxation time. Fig. 9 shows the DC conductivity data obtained from the above expression (Eq. 13) at various temperatures. The activation energy for the DC conductivity is calculated from the plot of ln ($\sigma_{DC}$) versus 1000/$T$ (Fig. 9) for these glasses. The plot is found to be linear and fitted using the following Arrhenius equation,

$$\sigma_{DC}(T) = B \exp\left( -E_{DC}/kT \right) \tag{14}$$

where $B$ is the pre-exponential factor, $E_{DC}$ is the activation energy for the DC conduction. The activation energy is calculated from the slope of the fitted line and found to be 2.58 ± 0.08 eV. This activation energy might be associated with the diffusion of non-bridging oxygen ions in the 648 -748 K temperature range.

The relaxation time associated with the above process is also determined from the plot of $M''$ versus frequency. The activation energy involved in the relaxation process of ions could be obtained from the temperature dependent relaxation time,

$$\tau_m = \tau_o \exp\left(\frac{E_R}{kT}\right) \qquad (15)$$

where $E_R$ is the activation energy associated with the relaxation process, $t_o$ is the pre-exponential factor, $k$ is the Boltzmann constant and $T$ is the absolute temperature. Fig. 9 shows a plot between ln ($t_m$) and 1000/$T$ along with the theoretical fit (solid line) to the above equation (Eq. 15). The value that is obtained for $E_R$ is 2.41±0.06 eV, which is in close agreement with that of the activation energy associated with dc conductivity. It suggests that similar energy barriers are involved in both the relaxation and conduction processes.

It is evident form Fig. 8 that the relaxation processes cannot be described in the whole frequency and temperature ranges using the electric modulus formalism. In this context, Scaling of modulus plots was proposed and found to be excellent technique to reveal the relaxation mechanism in the wide range of frequencies and temperatures. Dixon et.al. [17] have introduced a scaling formalism for the glasses. It is reported to be universal scaling and successfully employed to a variety of materials. According to this approach if one plots $w^{-1}$ log($f_{max}.M''/?M.f$) vs. $w^{-1}(1+w^{-1})$ log($f/f_{max}$), where $M''$ is the imaginary part of the electric modulus, *?M* is the electric modulus strength, *w* is the full width at half maximum of the peak of the imaginary part of electric modulus and $f_{max}$ is the frequency associated with the peak of the electric modulus. The plots at various temperatures are shown in Fig. 10. All the plots are successfully collapsed on the same curve indicating the validity of the above scaling on the glasses under investigation. Fig. 11 shows the normalized plots of electric modulus $M''$ versus frequency (*f*) wherein the frequency is scaled by the peak frequency. A perfect overlapping of all the curves on a single master curve is evidently noticed. This suggests that the conduction mechanism is temperature independent for the present glasses.

## 4. Conclusions:

Transparent glasses in the system $3BaO$-$3TiO_2$-$B_2O_3$ were fabricated via the conventional melt-quench technique. The dielectric data were analyzed using Jonscher's and electric modulus formalisms. The dielectric behavior was frequency independent especially at room temperature. DC conductivity was found to associate with these glasses at higher temperatures with the activation energy of 2.58 ± 0.08 eV which perhaps associated with oxygen ion diffusion. Dixon's scaling plots approach was found to be excellent for the entire temperature range under study. The relaxation dynamics were invariant of temperature.

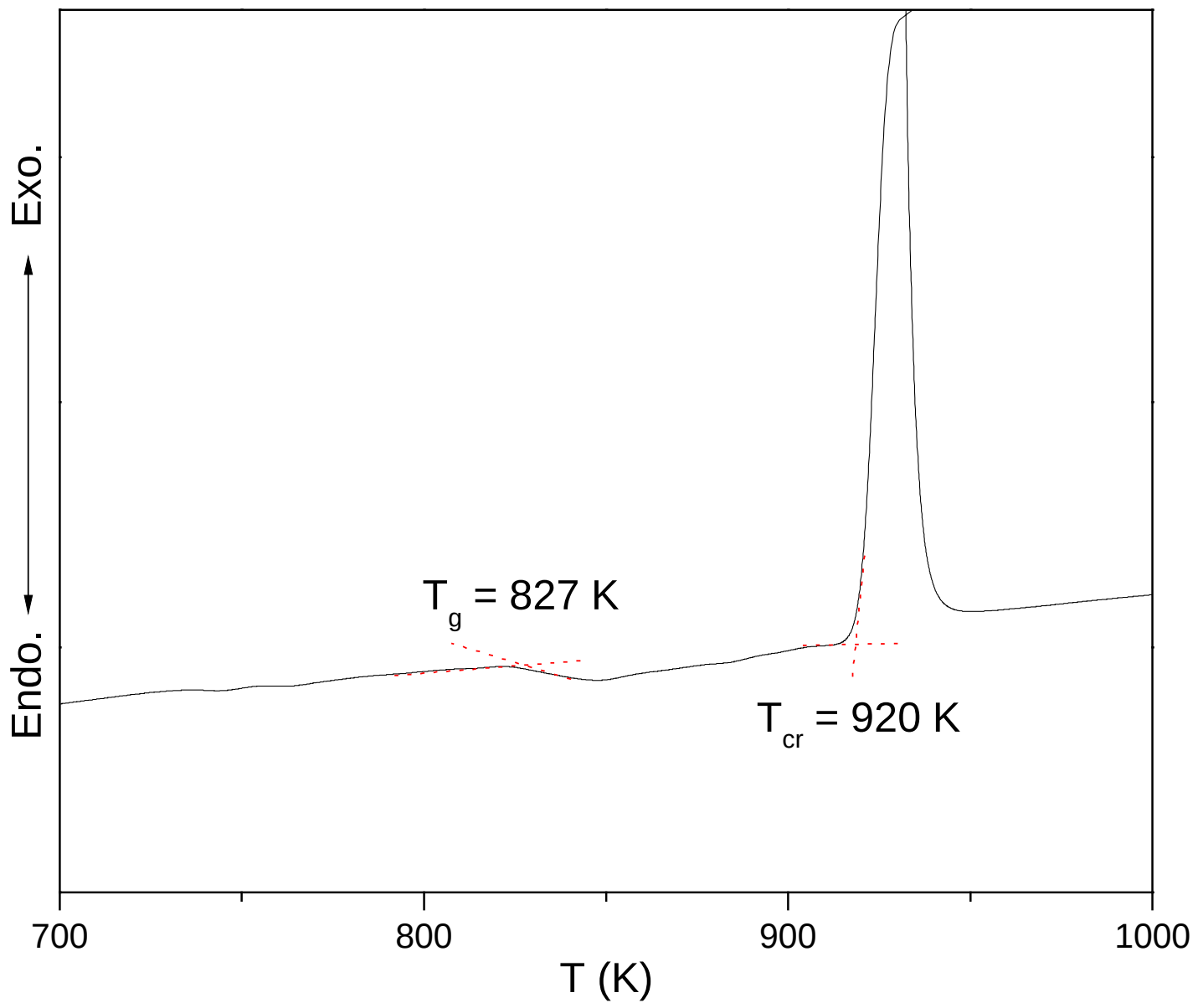

Fig. 1: DTA trace for the as-quenched $3BaO$-$3TiO_2$-$B_2O_3$ glass plate

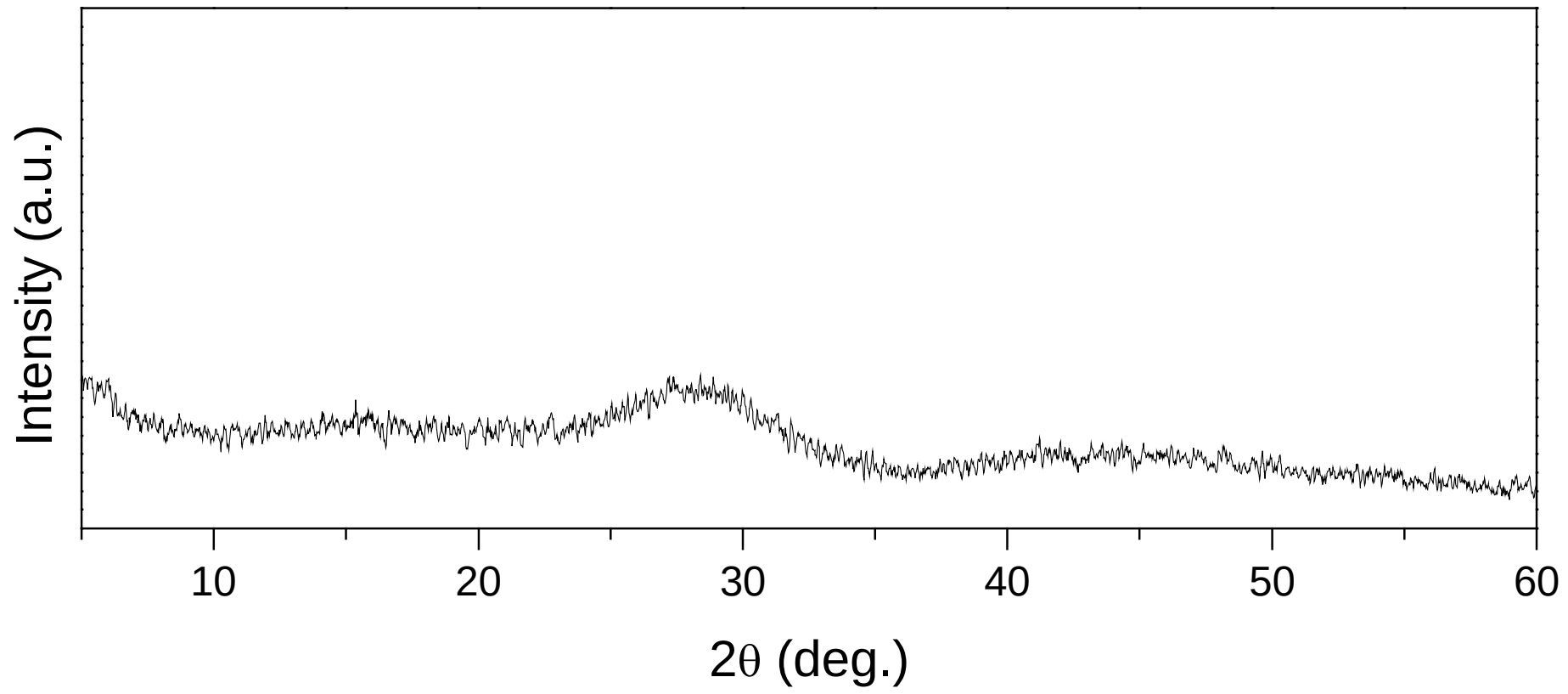

Fig. 2: X-ray diffraction pattern for the glass powder

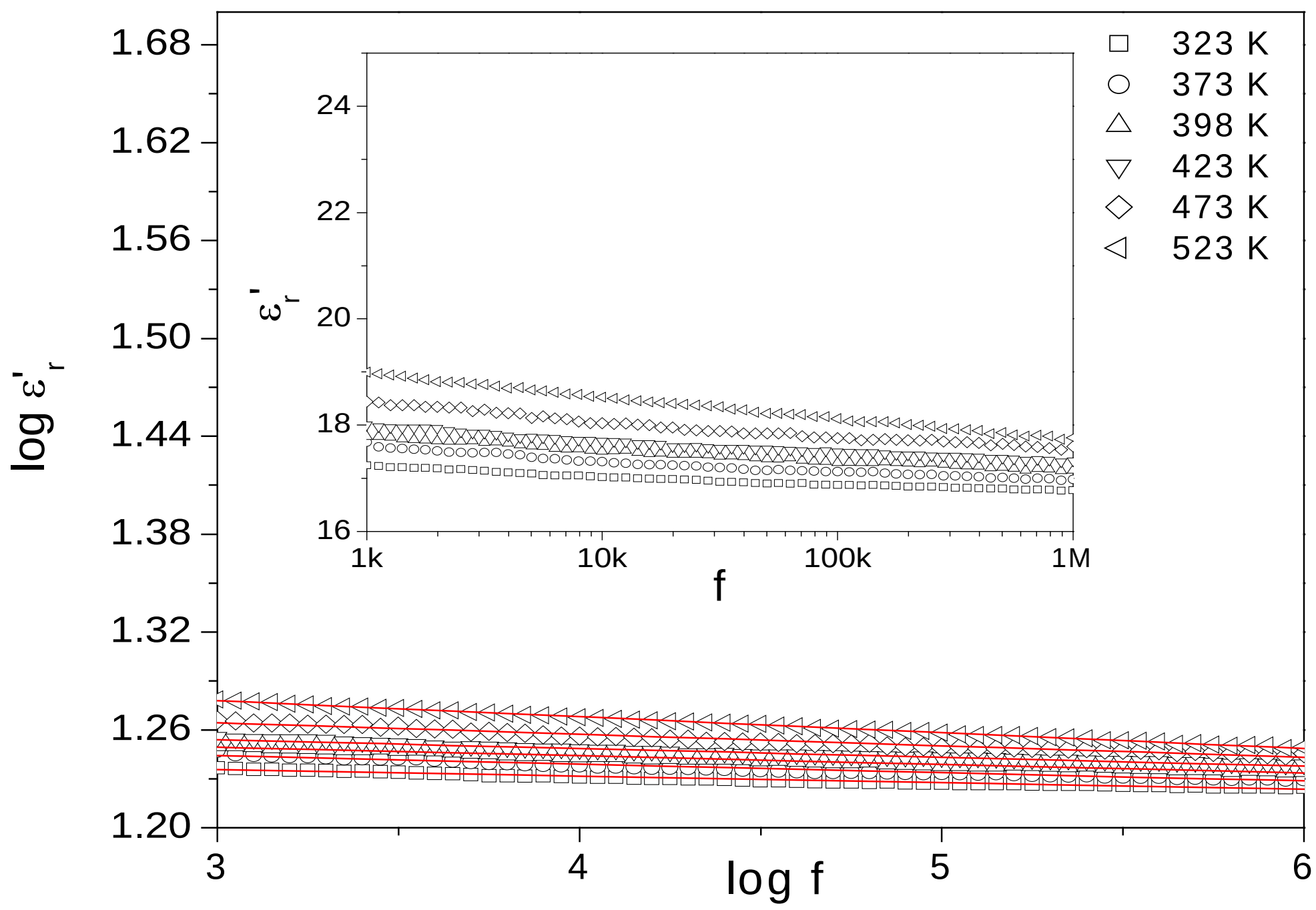

Fig. 3: log $e_r$' vs log f plots at various temperatures and the inset shows $e_r$' vs f plots

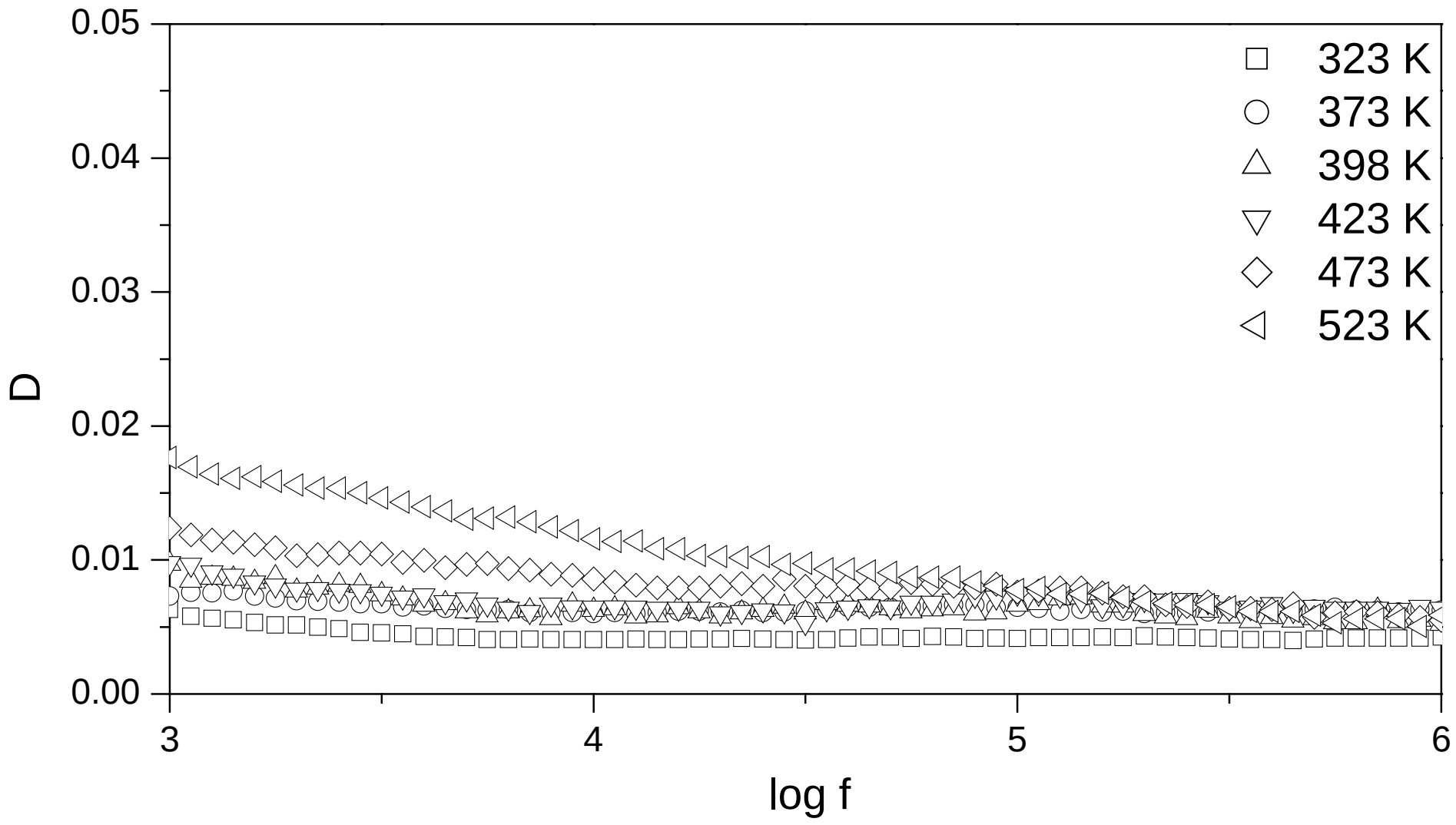

Fig. 4: D vs log f plots for the $3BaO-3TiO_2-B_2O_3$ glass-plates at various temperatures

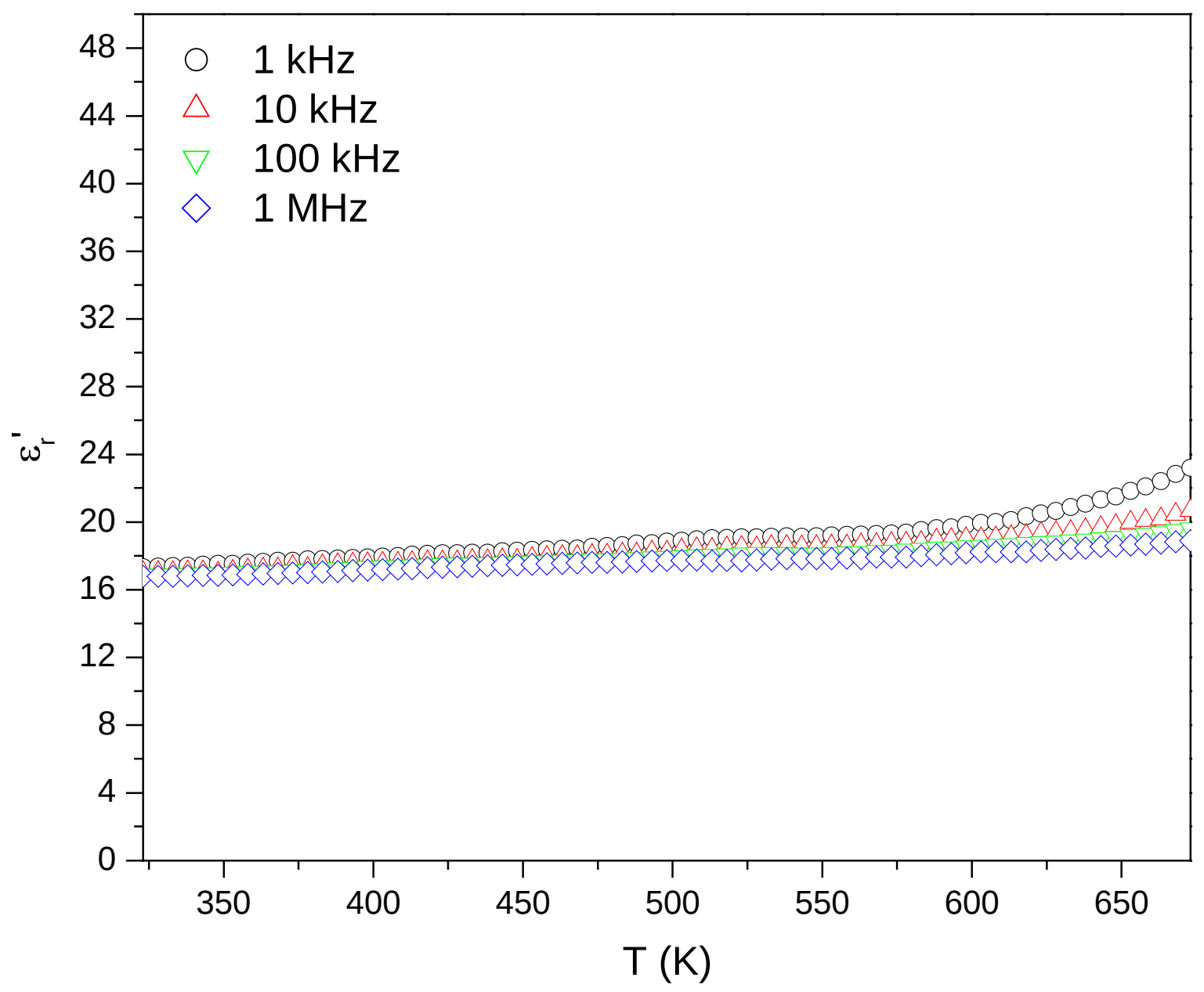

Fig. 5: Dielectric constant vs Temperature plots at four different frequencies

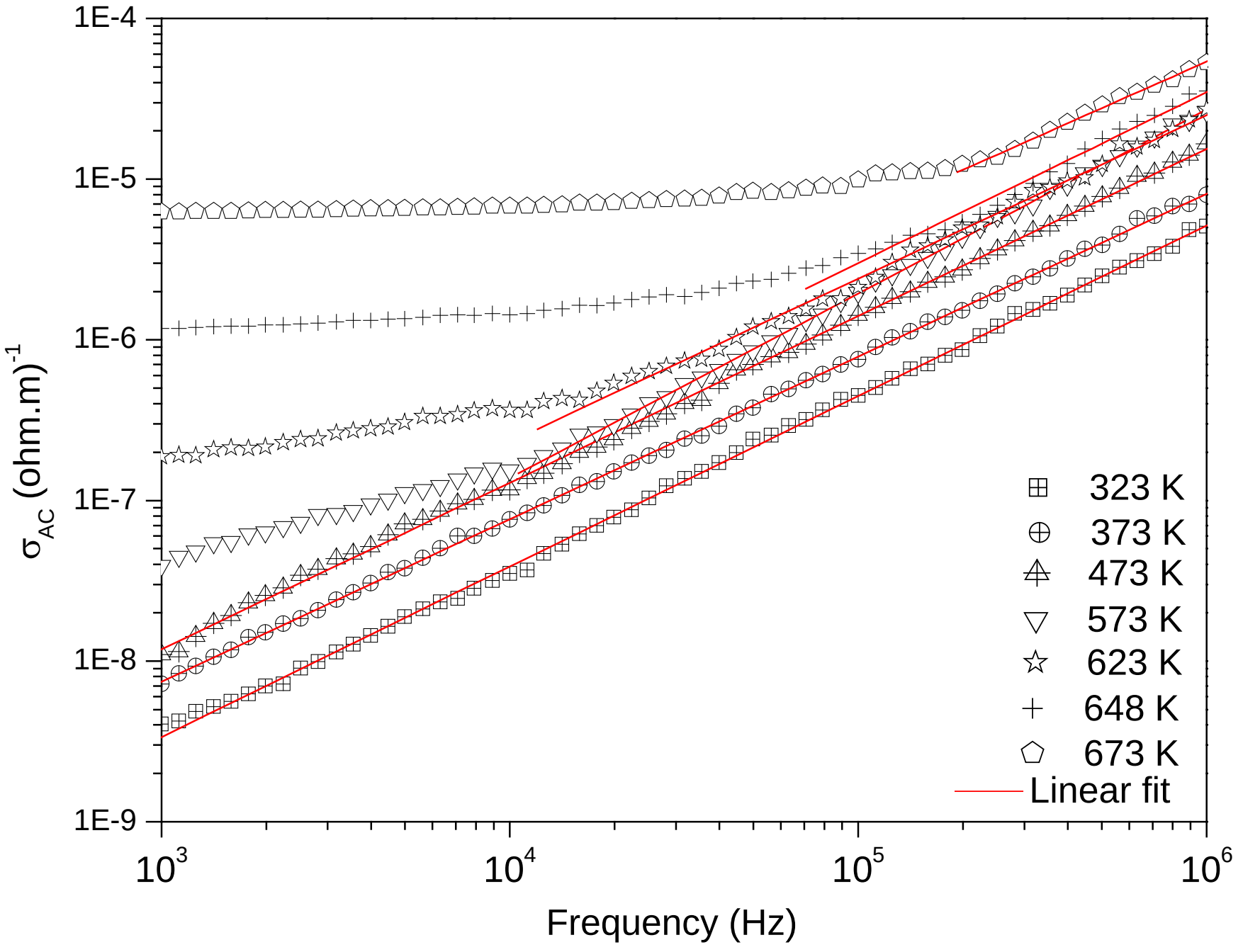

Fig. 6: Conductivity vs frequency plots at various temperatures

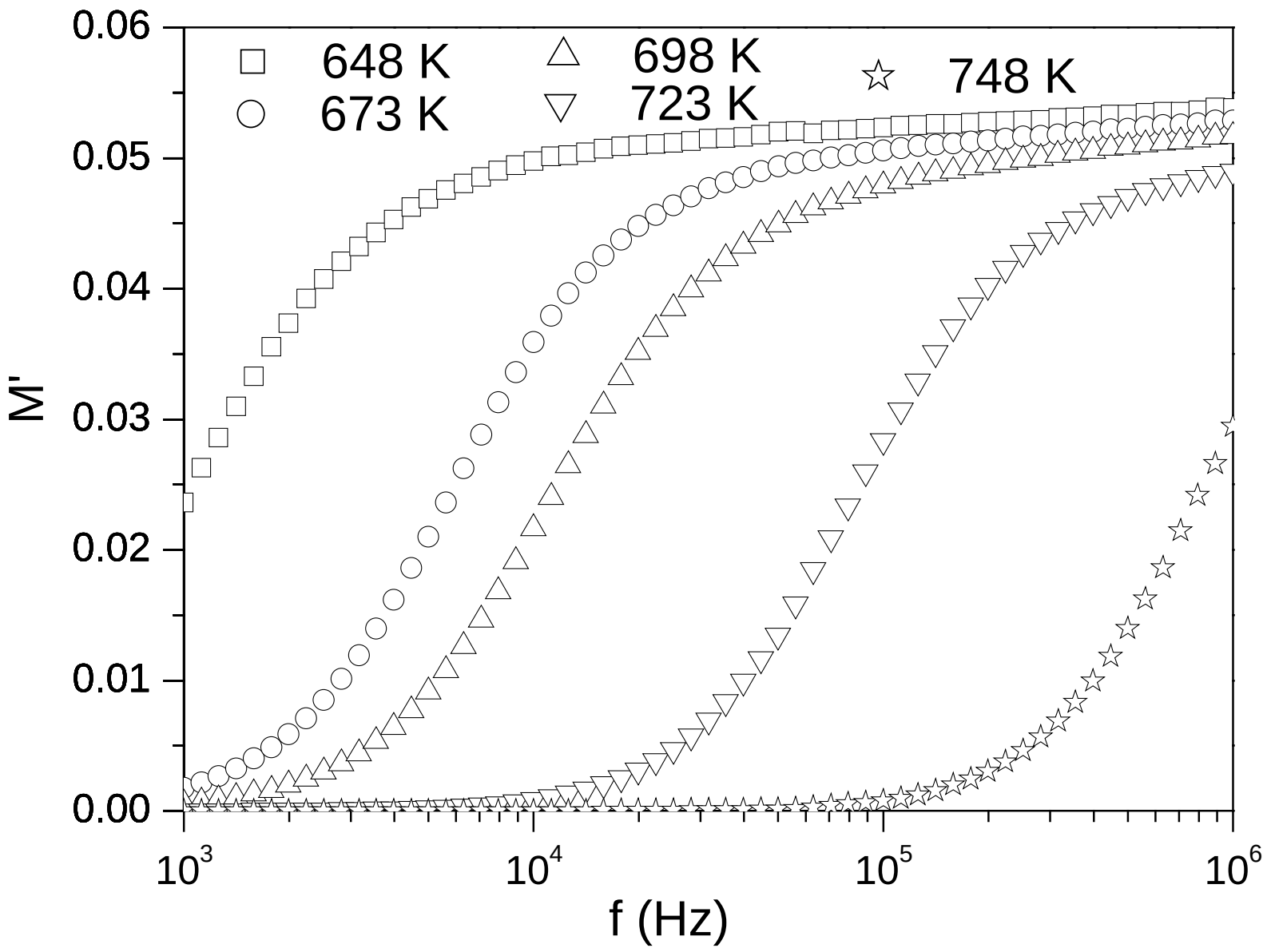

Fig. 7: Real part of the electric modulus as a function of frequency at different temperatures.

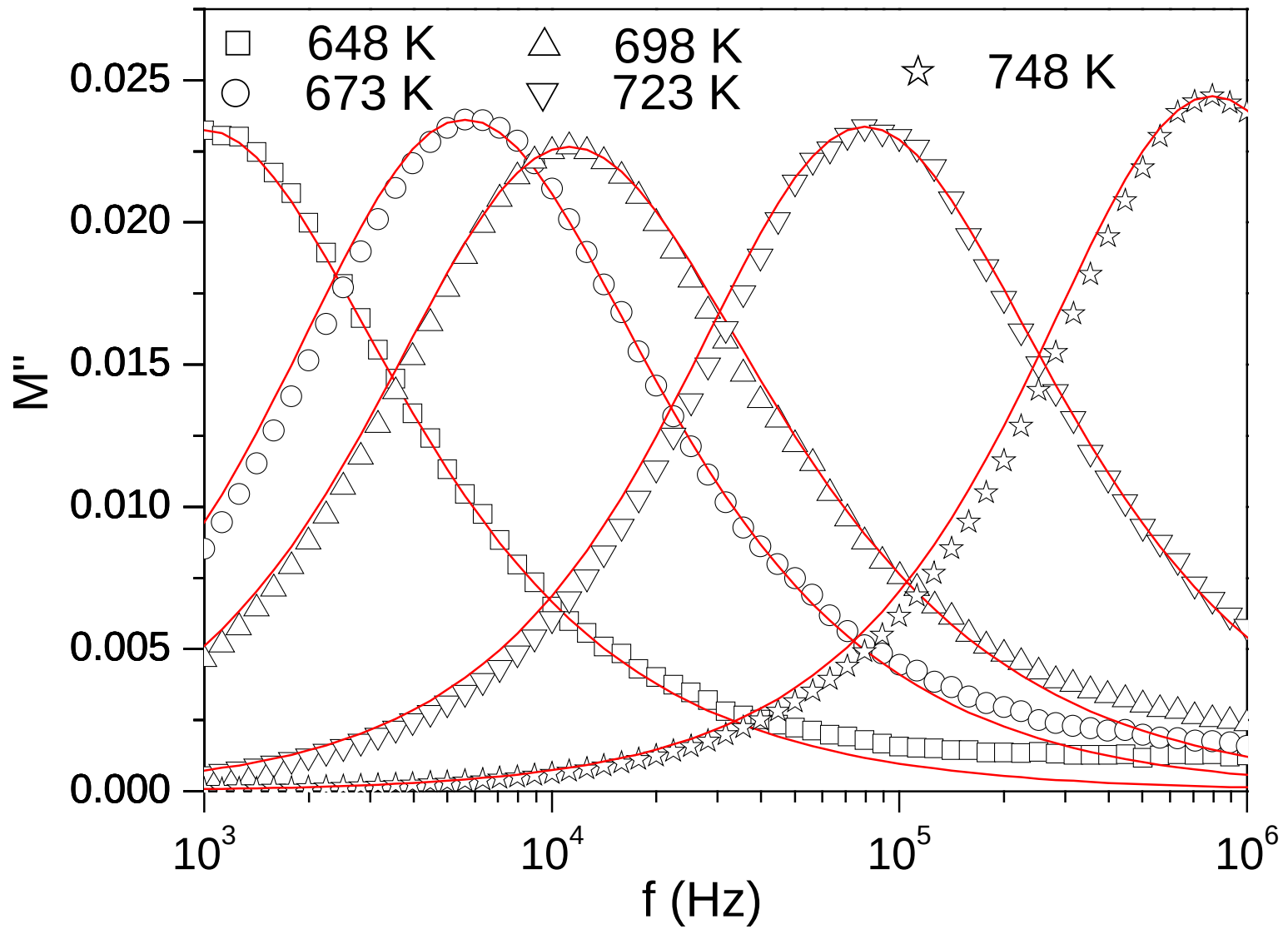

Fig. 8: Imaginary part of the electric modulus plots at various temperatures

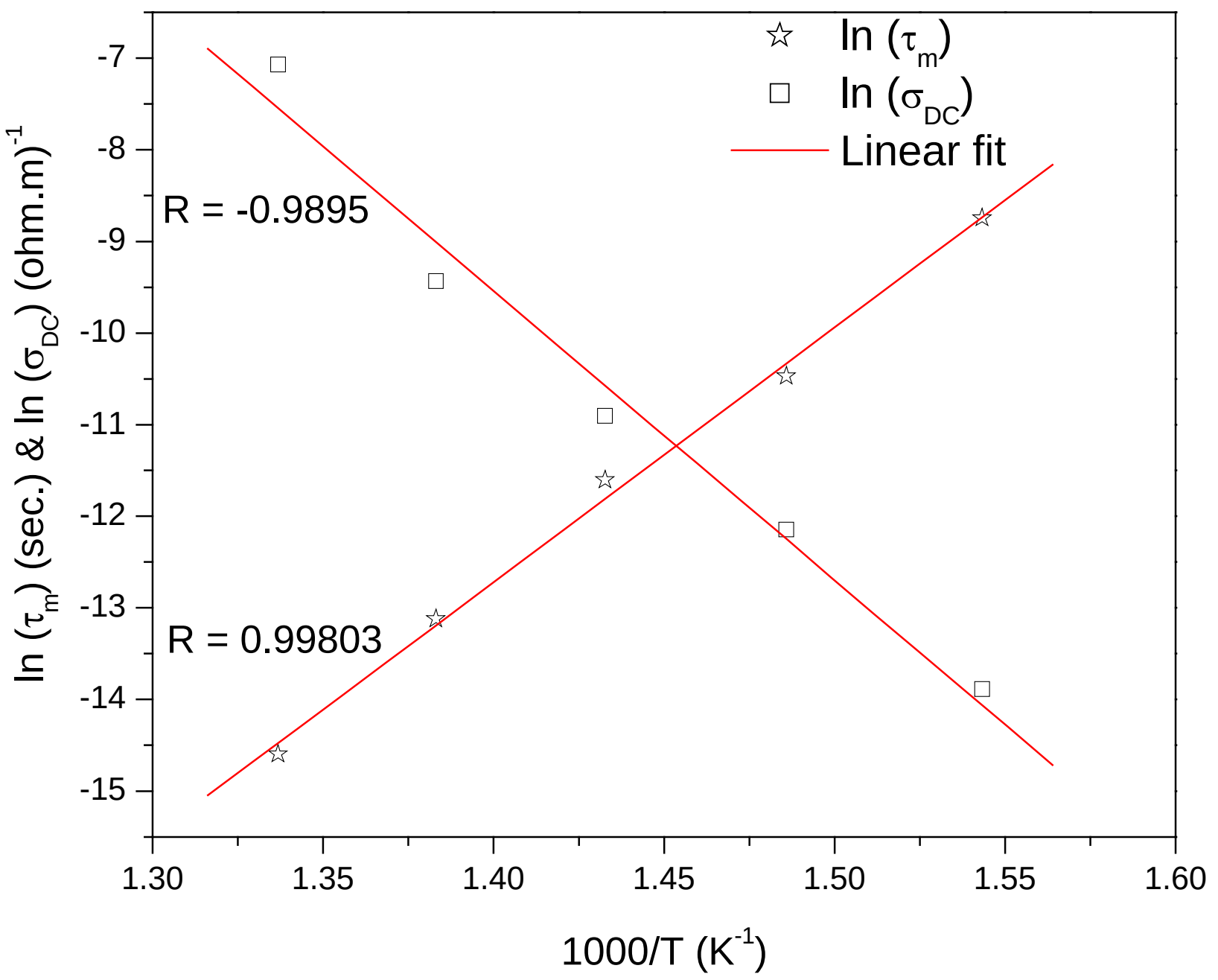

Fig. 9: Arrehenious plots for relaxation time and dc conductivity

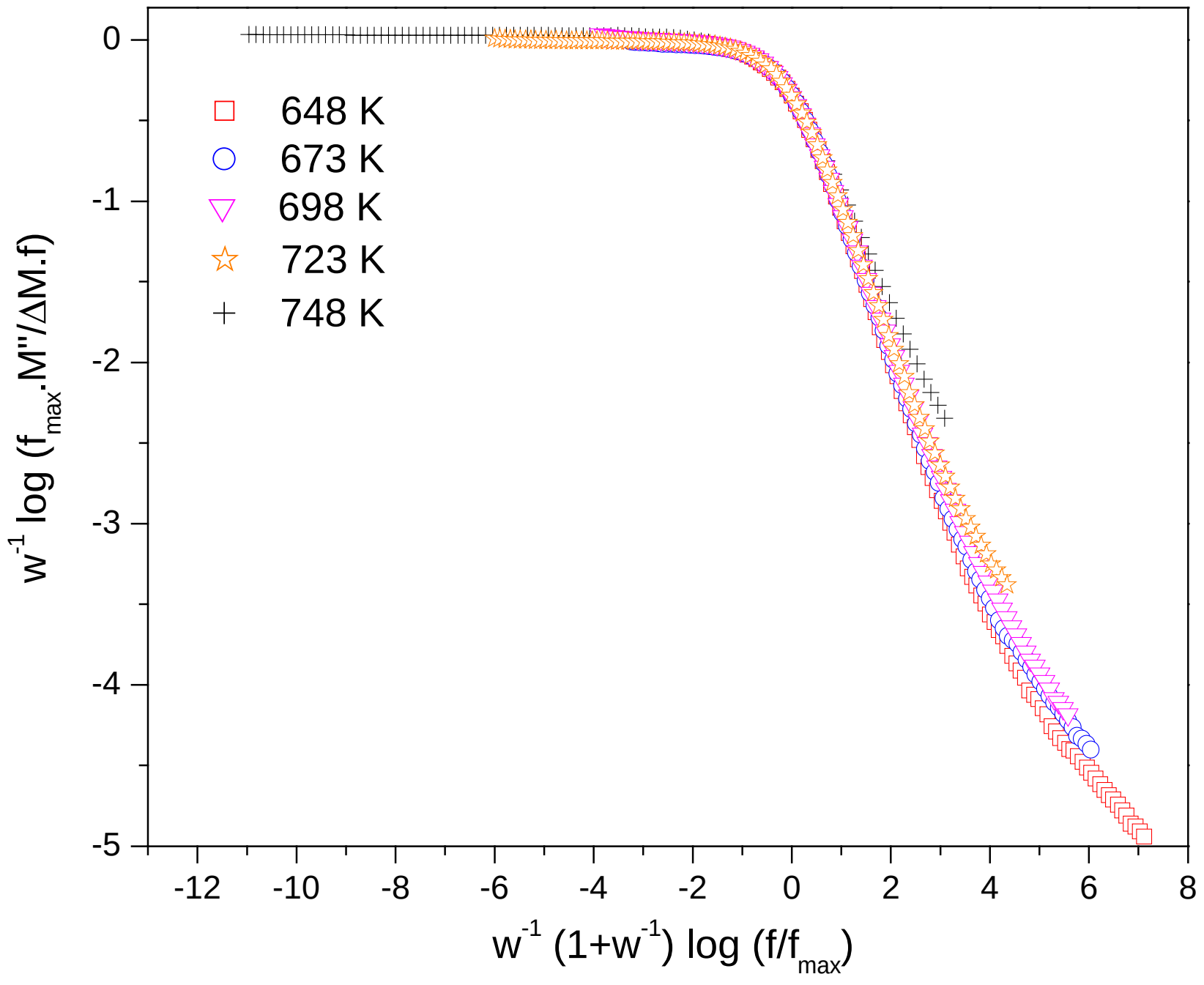

Fig. 10: Dixon-Nagel scaling plots for electric modulus data

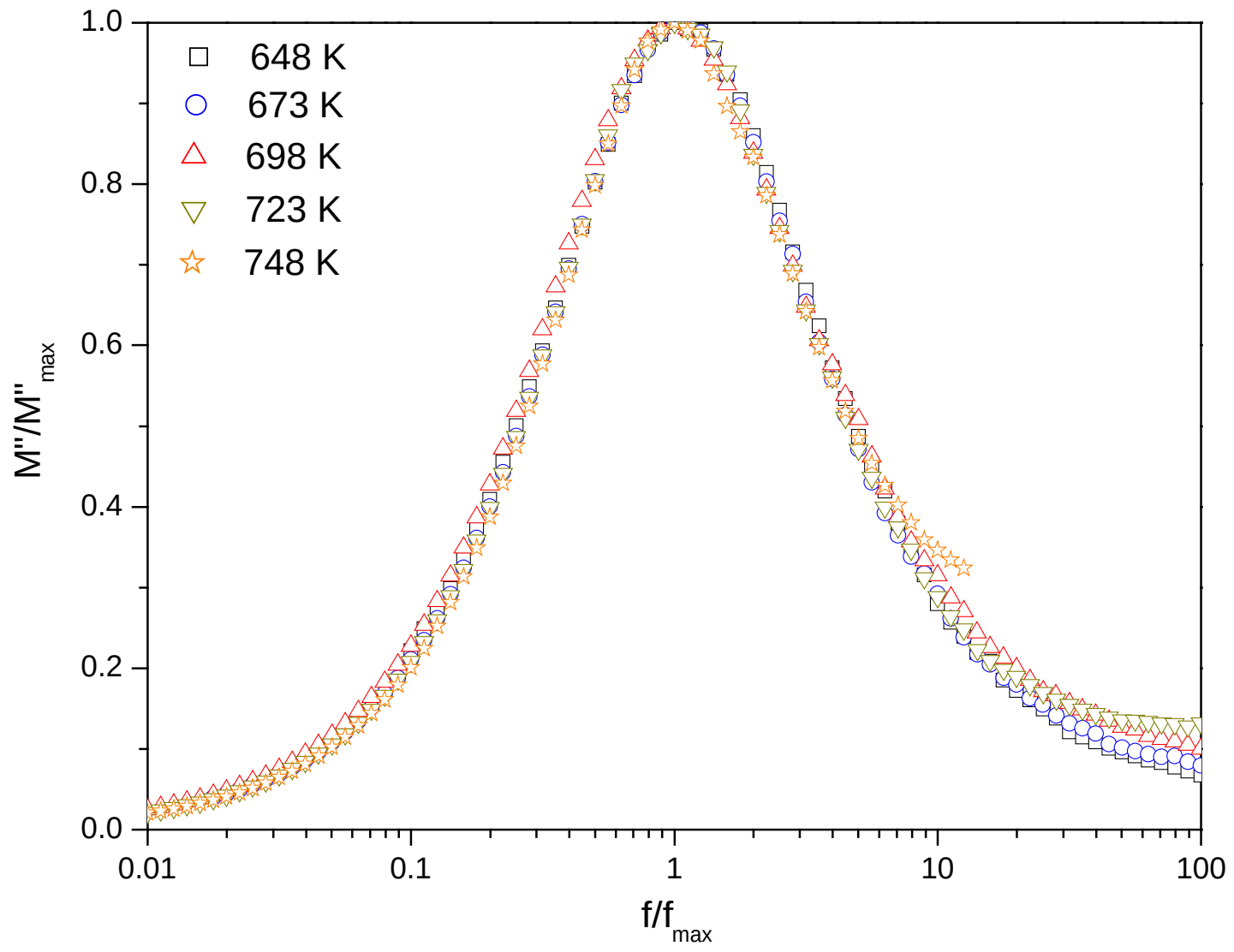

Fig. 11: Normalized plots of electric modulus vs normalized frequency at various temperatures